\documentstyle[aps,prd]{revtex}

\begin{document}
\title{Semiclassical Cosmological Perturbations Generated during
Inflation}
\author{Albert Roura and Enric Verdaguer \thanks{%
also at Institut de F\'\i sica d'Altes Energies (IFAE)}}
\address{Departament de F\'{\i}sica Fonamental,\\
Universitat de Barcelona, Av.~Diagonal 647,\\
08028 Barcelona, Spain}
\maketitle

\begin{abstract}
Environment interaction may induce stochastic semiclassical dynamics in
open quantum systems. In the gravitational context,
stress-energy fluctuations of quantum matter
fields give rise to a stochastic behaviour in the spacetime geometry.
Einstein-Langevin equation is a suitable tool to take these effects into
account when addressing the back-reaction problem in semiclassical
gravity.
We analyze within this framework the generation of gravitational
fluctuations during inflation, which are of great interest for large-scale
structure formation in cosmology.
\end{abstract}


\section{INTRODUCTION}

One of the key problems in modern cosmology is that of cosmic structure
formation \cite{padmanabhan93,kolb90}. If an inflationary period is
present, the initial seeds for
structure formation are supposed to be originated by the quantum
fluctuations of the inflaton field, which is responsible for driving
inflation \cite{linde90}. By semiclassical back reaction on the spacetime
geometry, these
quantum fluctuations will, in turn, produce fluctuations on the spacetime
metric. Here we want to look at this problem within the context of a
simple
chaotic inflationary model by means of a recently suggested formalism. In
this formalism classical metric fluctuations induced by quantum matter
fluctuations are described by a Langevin-type equation \cite{martin99}.
This is an
alternative to the more usual approach in which some perturbative degrees
of
freedom of the gravitational field are also quantized \cite{mukhanov92}.

The idea behind this approach is to relate the back-reaction problem in
semiclassical gravity with the dynamics of open quantum systems. In fact,
there are a number of situations in which one is interested in the
observables and the dynamics of a few degrees of freedom from a whole
closed
quantum system undergoing unitary evolution. These degrees of freedom
constitute an open system whose dynamics is no longer unitary due to its
interaction with the remaining degrees of freedom of the whole system,
which
constitute the environment \cite{feynman,paz93}.

For the existence of a semiclassical regime for the system dynamics two
requirements are needed \cite{omnes92,gellmann93}. The first is
decoherence,
which guarantees that
probabilities can be consistently assigned to histories describing the
evolution of the system. The second is that these probabilities should be
peaked near histories which correspond to solutions of classical equations
of motion. The effect of the environment plays a crucial role in the
semiclassical dynamics of the system. In fact, on the one hand, it may
provide enough induced decoherence through the entanglement between system
and environment \cite{zurek91,zurek93,paz93}. On the other hand, the
environment
back reaction on the
system dynamics will produce both dissipation and noise (commonly
connected
by fluctuation-dissipation relations). The environment may, thus, induce a
semiclassical stochastic dynamics on the system, which may be suitably
described by a Langevin-type equation \cite{gellmann93}.

The plan of the paper is the following: In Sec. \ref{sec2} we give a brief
summary of the Einstein-Langevin equation. We apply this formalism in Sec.
\ref{sec3} to
study the generation of cosmological gravitational perturbations during
inflation by considering the simplest model leading to chaotic inflation.
We
finally discuss our main conclusions in Sec. \ref{sec4}. Throughout the
paper
we use natural units ($\hbar =c=1$) and the $(+,+,+)$ sign convention of
Ref. \cite {misner73}.

\section{EINSTEIN-LANGEVIN EQUATION}

\label{sec2}

In the context of semiclassical gravity one treats the matter fields as
quantum fields on a classical curved spacetime. As a consequence of their
energy density, these fields act as gravitational sources which modify the
spacetime geometry. To study this back-reaction effect one usually uses
the
so-called semiclassical Einstein equation 
\begin{equation}
G_{ab}[g]=\frac{8\pi }{m_{p}^{2}}\langle T_{ab}[g,\hat{\phi}[g]]\rangle
_{ren},
\end{equation}
where the renormalized expectation value of the stress tensor operators of
the quantum matter fields in some quantum state are introduced as
gravitational sources. There are, however, some situations in which the
fluctuations of the stress tensor operator are important
\cite{kuo93/hu97}.
In those cases we cannot expect that the semiclassical Einstein equation
provides the actual dynamics of the spacetime metric any longer, but some
kind of {\em averaged} description.

It may be useful to consider the spacetime metric as an open system which
interacts gravitationally with the quantum matter fields, which constitute
the environment \cite{calzetta94,hu95/campos96/calzetta97}. In this case
the
system will exhibit a stochastic dynamics with fluctuations due to the
noise
induced by the environment. In order to take this effect into account, the
following modified equation, known as Einstein-Langevin equation, has been
suggested \cite{martin99}: 
\begin{equation}  \label{langevin}
G_{ab}[g+h]-\frac{8\pi}{m_p^2}\langle \hat{T}_{ab}[g+h]\rangle_{ren}=
\frac{%
8\pi}{m_p^2}\xi _{ab}[g],
\end{equation}
where $g$ is a solution of the semiclassical Einstein equation which is
used
as the background metric, whereas $h$ is a linear perturbation. The field
$%
\xi _{ab}[g]$ is a Gaussian stochastic classical source with the following
properties: 
\begin{eqnarray}
\langle \xi _{ab}(x)\rangle _{\xi } &=&0 \\
\langle \xi _{ab}(x)\xi _{cd}(y)\rangle _{\xi } &=&\frac{1}{2}\langle
\{\hat{%
t}_{ab}(x),\hat{t}_{cd}(y)\}\rangle[g] ,
\end{eqnarray}
where $\hat{t}_{ab}(x)\equiv \hat{T}_{ab}(x)-\langle\hat{T}_{ab}(x)
\rangle$%
. We use the two different notations $\langle \;\rangle _{\xi }$ and $%
\langle \;\rangle $ to explicitly distinguish the average associated to a
classical stochastic process from the expectation value of quantum
operators. The correlation function for the stochastic source, which will
generate a stochastic dynamics on the spacetime geometry, was precisely
chosen to take into account the quantum fluctuations of the stress tensor.

\section{COSMOLOGICAL PERTURBATIONS GENERATED DURING INFLATION}

\label{sec3}

Let us now consider the simplest model leading to chaotic inflation \cite
{linde90}, which is driven by a massive real scalar field $\hat{\phi}$
minimally coupled to the spacetime curvature (this field is usually called
the {\em inflaton}). The corresponding Lagrangian density is, thus: 
\begin{equation}
{\cal L}(\hat{\phi})=\frac{1}{2}g^{ab}\nabla _{a}\hat{\phi}\nabla
_{b}\hat{%
\phi}+\frac{1}{2}m^{2}\hat{\phi}^{2}
\end{equation}
A few comments are in order. First of all, the condition for the existence
of an inflationary period (characterized by an accelerated expansion of
spacetime) is that the value of the field averaged over a region with a
typical size equal to the Hubble radius (the so-called horizon scale) is
higher than the Planck mass, $m_{p}$. In fact, in order to have enough
inflation to solve the horizon and the flatness problem, more than 60
e-folds are needed. To achieve that, the scalar field should begin with a
value higher than $3 m_{p}$. On the other hand, as will be shown below,
the
small value of the CMB (Cosmic Microwave Background) large scale
anisotropies measured by COBE \cite{smoot92} imposes a severe constraint
on
the inflaton mass $m$, which should be of the order of $10^{-6}m_{p}$. 

We want to study small metric perturbations around a Robertson-Walker
geometry. For this purpose we need to deal with the corresponding gauge
freedom either by choosing a particular gauge or by working with gauge
invariant quantities \cite{mukhanov92}. We will restrict our study to
scalar-type perturbations of the metric. The expression for the perturbed
metric in the longitudinal gauge is then: 
\begin{equation}  \label{metric}
ds^{2}=a^{2}(\eta )\left( -(1+2\Phi (x))d\eta ^{2}+(1-2\Psi (x))\delta
_{ij}dx^{i}dx^{j}\right) ,
\end{equation}
where the two functions $\Phi (x)$ and $\Psi (x)$ correspond in this case
to
Bardeen's gauge invariant variables and $a^2(\eta)$ is the cosmological
scale factor of the background Robertson-Walker geometry. As shown below,
the Einstein-Langevin equation (\ref{langevin}) is gauge invariant.
Therefore, we can work in a given gauge and finally extract the desired
gauge invariant quantities in a consistent way. To see how the first
member
of Eq. (\ref{langevin}) is gauge invariant, one uses the following result
for linear perturbations in $h$:

\begin{quote}
$A_{b}^{a}[g+h]$ is gauge invariant if and only if ${\cal L}_{\vec
{\varsigma%
}}(A_{b}^{a}[g])=0$ for any vector field $\vec{\varsigma} (x)$ and this is
equivalent to $A_{b}^{a}[g]\propto\delta _{b}^{a}$ (zero being a
particular
case).
\end{quote}

The first member of our Einstein-Langevin equation is, thus, gauge
invariant
if $G_{ab}^{(0)}[g]-(8\pi /m_{p}^{2})\ \langle
\hat{T}_{ab}^{(0)}[g]\rangle
_{ren}=0$, but this is indeed the case since the background metric $g$ is
taken to be a solution of the semiclassical Einstein equation. On the
other
hand, the second member of Eq. (\ref{langevin}) is explicitly gauge
invariant since it does not depend on the perturbed metric.

It is convenient to decompose the inflaton scalar field in the following
way: $\hat{\phi}(x)=\phi (t)+\hat{\varphi}(x)$, where $\phi (t)$ is the
homogeneous background solution, which is compatible with the background
metric through the semiclassical Einstein equation, whereas
$\hat{\varphi}%
(x) $ corresponds to a free massive quantum scalar field with zero
expectation value on the spacetime with the background metric: $\langle
\hat{%
\varphi}(x)\rangle _{g}=0$. The two main ingredients that we need for our
Einstein-Langevin equation are the renormalized expectation value of the
stress tensor on the spacetime with the perturbed metric $\tilde{g}=g+h$,
and the noise kernel, which takes into account the fluctuations of the
stress tensor evaluated on the background metric. The stress tensor of a
minimally coupled massive scalar field is: 
\begin{equation}
\hat{T}_{\mu \nu }=\partial _{\mu }\hat{\phi}\partial _{\nu }\hat{\phi}+%
\frac{1}{2}\tilde{g}_{\mu \nu }(\partial _{\mu }\hat{\phi}\partial ^{\mu
}%
\hat{\phi}+m^{2}\hat{\phi}^{2}) .
\end{equation}
Using the decomposition for the scalar field introduced above, we rewrite
the renormalized expectation value for the stress tensor as 
\begin{equation}  \label{stress}
\langle \hat{T}_{\mu \nu }[g+h]\rangle ^{ren}=\langle \hat{T}_{\mu \nu
}[g+h]\rangle _{\phi \phi }+\langle \hat{T}_{\mu \nu }[g+h]\rangle _{\phi
\varphi }+\langle \hat{T}_{\mu \nu }[g+h]\rangle _{\varphi \varphi }^{ren}
,
\end{equation}
where only the homogeneous solution for the scalar field contributes to
the
first term. The second term is proportional to $\langle \hat{\varphi}%
[g+h]\rangle $, but this quantity is no longer zero since the field
dynamics
is considered on the perturbed spacetime. Finally, the last term
corresponds
to the expectation value of the stress tensor for a free scalar field on a
spacetime with the perturbed metric. In the usual approach when computing
fluctuations during inflation, $\hat{\varphi}$ is treated perturbatively.
This last term being quadratic in $\hat{\varphi}$, is of higher order and
will not be taken into account.

As for the noise kernel, after using the previous decomposition, the
following expression is obtained: 
\begin{equation}
\langle \{\hat{t}_{\mu \nu },\hat{t}_{\rho \sigma }\}\rangle [g]=\langle
\{%
\hat{t}_{\mu \nu },\hat{t}_{\rho \sigma }\}\rangle _{\phi \varphi
}[g]+\langle \{\hat{t}_{\mu \nu },\hat{t}_{\rho \sigma }\}\rangle
_{\varphi\varphi }[g],
\end{equation}
where we have used the fact that $\langle \hat{\varphi}\rangle _{g}=0=$ $%
\langle \hat{\varphi}\hat{\varphi}\hat{\varphi}\rangle _{g}$ for Gaussian
states (those considered here) on the background geometry. It is important
to note that both contributions to the noise kernel (the first term is
quadratic in $\hat{\varphi}$ whereas the second one is quartic) are
``conserved'' separately since both $\phi (t)$ and $\hat{\varphi}(x)$
satisfy the Klein-Gordon equation on the background geometry. Due to this
fact, the two corresponding stochastic sources can be consistently
considered in an independent way. We are, thus, allowed to concentrate on
the source associated to the first term from now on. The contribution of a
term of the same sort as the second one has been discussed elsewhere \cite
{roura99b}. One can check that the {\em space-space} components coming
from
the stress-tensor expectation value terms that we are considering and the
stochastic source are diagonal, {\it i.e.}, $\langle \hat{T}_{ij}\rangle
=0=\xi _{ij}$ for $i\neq j$. This, in turn, implies that the two gauge
invariant quantities used to characterize the scalar-type metric
perturbations must be equal: $\Phi =\Psi$ \cite{mukhanov92}.

Let us write the Einstein-Langevin equation in Fourier space and consider
the $0i$-component: 
\begin{equation}
2ik_{i}({\cal H}\Phi _{k}+\Phi _{k}^{\prime })=\frac{8\pi }{m_{p}^{2}}\xi
_{k\;0i},  \label{0i}
\end{equation}
where $k_{i}$ is the comoving momentum component associated to the
comoving
coordinate $x^{i}$ (throughout the paper we use the subindex $k$ to denote
the comoving momentum vector $\vec{k}$ that labels the Fourier modes in
flat
space), primes denote derivatives with respect to the conformal time $\eta
$
and ${\cal H}\equiv a^{\prime }(\eta )/a(\eta )$. The first member is just
the linearized Einstein tensor for the perturbed metric (\ref{metric})
\cite
{mukhanov92}. There should also appear a non-local term of {\em
dissipative}
character coming from the second term in (\ref{stress}), which we have not
considered in this work, where we are mainly concerned about the
fluctuating
part.

From this equation we may obtain the metric perturbations $\Phi _{k}$ in
terms of the stochastic source $\xi _{k\;0i}$. For this purpose we need
the
retarded propagator for the gravitational potential $\Phi _{k}$, {\it
i.e.},
the required Green function to solve the inhomogeneous first order
differential Eq. (\ref{0i}) with the appropriate boundary conditions: 
\begin{equation}
\tilde{G}_{k}^{ret}(\eta ,\eta ^{\prime })=-i\frac{4\pi }{k_{i}m_{p}^{2}}%
\left( \theta (\eta -\eta ^{\prime })\frac{a(\eta ^{\prime })}{a(\eta )}%
+f(\eta ,\eta ^{\prime })\right) ,
\end{equation}
where $f(\eta ,\eta ^{\prime })$ is a homogeneous solution related to the
chosen initial conditions. If we take, for instance, $f(\eta ,\eta
^{\prime
})=-\theta (\eta _{0}-\eta ^{\prime })\ a(\eta ^{\prime })/a(\eta )$, we
would obtain the stochastic evolution of the metric perturbations for
$\eta
>\eta _{0}$ due to the effect of the stochastic source after $\eta _{0}$.
The correlation function for the metric perturbations is then given by the
following expression: 
\begin{eqnarray}
\langle \Phi _{k}(\eta )\Phi _{k^{\prime }}(\eta ^{\prime })\rangle _{\xi
}
&=&(2\pi )^{3}\delta (\vec{k}+\vec{k}^{\prime })\int^{\eta }d\eta
_{1}\int^{\eta ^{\prime }}d\eta _{2}\tilde{G}_{k}^{ret}(\eta ,\eta _{1})%
\tilde{G}_{k^{\prime }}^{ret}(\eta ^{\prime },\eta _{2})  \nonumber \\
&&\cdot \langle \xi _{k\;0i}(\eta _{1})\xi _{k^{\prime }\;0i}(\eta
_{2})\rangle _{\xi }.
\end{eqnarray}
And the correlation function for the stochastic source is, in turn,
connected with the stress-energy fluctuations: 
\begin{equation}
\langle \xi _{k\;0i}(\eta _{1})\xi _{-k\;0i}(\eta _{2})\rangle _{\xi
}=\frac{%
1}{2}\langle \{\hat{t}_{0i}^{k}(\eta _{1}),\hat{t}_{0i}^{-k}(\eta
_{2})\}\rangle _{\phi \varphi }=\frac{1}{2}k_{i}k_{i}\phi ^{\prime }(\eta
_{1})\phi ^{\prime }(\eta _{2})G_{k}^{(1)}(\eta _{1},\eta _{2}),
\end{equation}
where $G_{k}^{(1)}(\eta _{1},\eta _{2})=\langle \{\hat{\varphi}_{k}(\eta
_{1}),\hat{\varphi}_{-k}(\eta _{2})\}\rangle $ is the $k$-mode Hadamard
function for a free minimally coupled scalar field which is in a state
close
to the Euclidean vacuum on an almost de Sitter background.

The so-called ``slow-roll'' parameters account for the fact that the
background geometry is not exactly that of de Sitter spacetime (for which
$%
a(\eta )=-1/H\eta $ with $-\infty <\eta <0$). It is also useful to compute
the Hadamard function for a massless field and consider a perturbative
expansion in terms of the dimensionless parameter $m/m_{p}$, for which
observations seem to imply, as will be seen below, a value of the order of
$%
10^{-6}$. Thus, we will consider $\bar{G}_{k}^{(1)}(\eta _{1},\eta
_{2})=a(\eta _{1})a(\eta _{2})
G_{k}^{(1)}(\eta _{1},\eta _{2})=\langle 0|\{\hat{y}_{k}(\eta_{1}),
\hat{y}_{-k}(\eta _{2})\}|0\rangle$ such that $\hat{a}_k|0\rangle=0$ with
$\hat{y}_{k}(\eta)=a(\eta)\hat{\varphi}_{k}(\eta)=
\hat{a}_k u_k(\eta)+\hat{a}^\dagger_{-k} u_{-k}^{*}(\eta)$
and $u_k(\eta)=(2k)^{-1/2}e^{-ik\eta}(1-i/k\eta)$
corresponding to the positive frequency $k$-mode for a massless minimally
coupled
scalar field in the Euclidean vacuum state on a de Sitter background
\cite{birrell84}.

The result to lowest order in the mass $m$ of the inflaton field and the
``slow-roll'' parameters is: 
\begin{eqnarray}
\langle \Phi _{k}(\eta )\Phi _{k^{\prime }}(\eta ^{\prime })\rangle _{\xi
}
&=&\frac{64\pi ^{5}}{m_{p}^{4}}\left( a(\eta )a(\eta ^{\prime })\right)
^{-1}\delta (\vec{k}+\vec{k}^{\prime })\int_{\eta _{0}}^{\eta }d\eta
_{1}\int_{\eta _{0}}^{\eta ^{\prime }}d\eta _{2}a(\eta _{1})a(\eta
_{2})\dot{%
\phi}(\eta _{1})\dot{\phi}(\eta _{2})\bar{G}_{k}^{(1)}(\eta _{1},\eta
_{2}) 
\nonumber \\
&=&64\pi ^{5}\left( \frac{m}{m_{p}}\right) ^{2}k^{-3}\delta
(\vec{k}+\vec{k}%
^{\prime })\int_{k\eta _{0}}^{k\eta }d(k\eta _{1})\int_{k\eta _{0}}^{k\eta
^{\prime }}d(k\eta _{2})\frac{k\eta }{k\eta _{1}}\frac{k\eta ^{\prime }}{%
k\eta _{2}}  \nonumber \\
&&\cdot \left[ \cos k(\eta _{1}-\eta _{2})\cdot \left( 1+\frac{1}{k\eta
_{1}k\eta _{2}}\right) -\sin k(\eta _{1}-\eta _{2})\cdot \left( \frac{1}{%
k\eta _{1}}-\frac{1}{k\eta _{2}}\right) \right]   \nonumber \\
&=&64\pi ^{5}\left( \frac{m}{m_{p}}\right) ^{2}k^{-3}\delta
(\vec{k}+\vec{k}%
^{\prime })\left[ \cos k(\eta -\eta ^{\prime })-\frac{1}{k\eta _{0}}\left(
k\eta \cos k(\eta -\eta _{0})\right. \right.   \nonumber \\
&&\left. \left. +k\eta ^{\prime }\cos k(\eta ^{\prime }-\eta _{0})\right)
+%
\frac{k\eta k\eta ^{\prime }}{(k\eta _{0})^{2}}\right] \text{,}
\label{power1}
\end{eqnarray}
where we used the lowest order approximation for $\dot{\phi}(t)$ during
``slow-roll'' (overdots denote derivatives with respect to the physical
time 
$t$): $\dot{\phi}(t)\simeq -m_{p}^{2}(m/m_{p})$. We considered the effect
of
the stochastic source after the conformal time $\eta _{0}$. Notice that
the
result (\ref{power1}) is rather independent of the value of $\eta _{0}$
provided that it is negative enough,{\em \ i.e.}, it corresonds to an
early
enough initial time. This weak dependence on the initial conditions is
rather usual in this context and can be qualitatively understood: after a
sufficient amount of time, the accelerated expansion for the quasi-de
Sitter
spacetime during inflation effectively erases any information about the
initial conditions, which is redshifted away. The actual result will,
therefore, be very close to that for $\eta _{0}=-\infty $: 
\begin{equation}
\langle \Phi _{k}(\eta )\Phi _{k^{\prime }}(\eta ^{\prime })\rangle _{\xi
}=8\pi ^{2}\left( \frac{m}{m_{p}}\right) ^{2}k^{-3}(2\pi )^{3}\delta
(\vec{k}%
+\vec{k}^{\prime })\cos k(\eta -\eta ^{\prime })\text{.}  \label{final}
\end{equation}

\section{CONCLUSIONS}

\label{sec4}It is of major interest to study the cosmological implications
which can be extracted from our work, especially those related to
large-scale gravitational fluctuations. These fluctuations are believed to
play a crucial role in the generation of the large-scale structure and
matter distribution observed in our present universe \cite{padmanabhan93}.
They are also tightly connected with the anisotropies in the CMB
radiation,
which decoupled from matter about $3\cdot 10^{5}$ years after the Big Bang
and provides us with very valuable information about the early universe
\cite
{kolb90}.

From the analysis of our final result in Eq.\ (\ref{final})
two main facts can be concluded. First, an almost Harrison-Zel'dovich
scale-invariant spectrum seems to be obtained for large scales (small
values
of $k$). Second, no significant relaxation of the coupling parameter is
found. Since we get $\langle \Phi _{k}(\eta )\Phi _{k^{\prime }}(\eta
^{\prime })\rangle _{\xi }\propto (m/m_{p})^{2}$ in agreement with the
usual
results \cite{mukhanov92,tanaka98}, the small value of the CMB anisotropies
detected
by COBE imposes a severe bound on the gravitational fluctuations,
characterized by $\langle \Phi _{k}(\eta )\Phi _{k^{\prime }}(\eta
^{\prime
})\rangle _{\xi }$, which implies $(m/m_{p})\sim 10^{-6}$, whereas the
mechanisms considered in those works
\cite{calzetta95/matacz97/gonorazky97}
which allowed an important relaxation of this fine tuning (due to the
extremely homogeneous classical initial conditions taken for the inflaton
field) resulted in $\langle \Phi _{k}(\eta )\Phi _{k^{\prime }}(\eta
^{\prime })\rangle _{\xi }\propto (m/m_{p})$.

It can be shown that genuine quantum correlation functions can be
equivalently obtained through a stochastic description based on
Langevin-type equations even in regimes where the actual dynamics of the
system does not admit a description in classical terms \cite{calzetta00}.
The case of gravitational perturbations coupled to a scalar field is more
subtle due to the existing gauge symmetry associated to diffeomorphic
transformations and the subsequent constraints arising in the dynamics of
the whole system. Nevertheless, total agreement with the purely quantum
treatment \cite{mukhanov92} is expected at least for the case in which
both
gravitational inhomogenities and the scalar field are treated
perturbatively
to linear order \cite{roura00}.

\section*{ACKNOWLEDGEMENTS}

We are grateful to Bei-Lok Hu, Esteban Calzetta and Rosario Mart\'\i n for
interesting discussions. This work has been partially supported by the
CICYT
Research Project No. AEN98-0431. A.\ R.\ also acknowledges support of a
grant from the Generalitat de Catalunya.

\end{document}